\documentclass[a4paper,referee]{raa}     

\usepackage{graphicx,times}
\usepackage{natbib}
\usepackage{amssymb,amsmath}
\usepackage{threeparttable}
\usepackage{upgreek}

\makeatletter

\newcommand{\Rmnum}[1]{\expandafter\@slowromancap\romannumeral #1@}
\makeatother

\defcitealias{2015AJ....149..199D}{Du15}
\defcitealias{2018MNRAS.476.4488H}{H18}
\defcitealias{2023MNRAS.520L..85S}{S23}

\begin{document}

\title{Low Surface Brightness Galaxies selected by different model fitting}
\volnopage{ {\bf 20XX} Vol.\ {\bf X} No. {\bf XX}, 000--000}
\setcounter{page}{1}

\author{Bing-qing Zhang\inst{1,2}, Hong Wu\inst{1,2}, Wei Du\inst{1}, Pin-song Zhao\inst{1}, Min He\inst{1}, Feng-jie Lei\inst{1}}

%\author{Bing-qing Zhang\inst{1,2}~\orcidlink{0000-0002-6659-1152}, Hong Wu\inst{1,2}~\orcidlink{0000-0002-4333-3994}, Wei Du\inst{1}~\orcidlink{0000-0003-4546-8216}, Pin-song Zhao\inst{1}~\orcidlink{0000-0002-4328-538X}, Min He\inst{1}~\orcidlink{0000-0001-6139-7660}, Feng-jie Lei\inst{1}~\orcidlink{0000-0003-0408-5497}}

\institute{Key Laboratory of Optical Astronomy, National Astronomical Observatories, Chinese Academy of Sciences, 20A Datun Road, Chaoyang District, Beijing, 100101, China; {\it hwu@bao.ac.cn, bqzhang@bao.ac.cn}\\
\and School of Astronomy and Space Science, University of Chinese Academy of Sciences, No. 19A Yuquan Road, Beijing, 100049, China \\
%\vs \no
{\small accepted 2023 November 2nd}}

\abstract{
We present a study of low surface brightness galaxies (LSBGs) selected by fitting the images for all the galaxies in $\upalpha$.40 SDSS DR7 sample with two kinds of single-component models and two kinds of two-component models (disk+bulge): single exponential, single s\'{e}rsic, exponential+deVaucular (exp+deV), and exponential+s\'{e}rsic (exp+ser).
Under the criteria of the B band disk central surface brightness $\mu_{\rm 0,disk}{\rm (B) \geqslant 22.5\ mag\ arcsec^{-2}}$ and the axis ratio $\rm b/a > 0.3$, we selected four none-edge-on LSBG samples from each of the models which contain 1105, 1038, 207, and 75 galaxies, respectively. 
There are 756 galaxies in common between LSBGs selected by exponential and s\'{e}rsic models, corresponding to 68.42\% of LSBGs selected by the exponential model and 72.83\% of LSBGs selected by the s\'{e}rsic model, the rest of the discrepancy is due to the difference in obtaining $\mu_{0}$ between the exponential and s\'{e}rsic models. Based on the fitting, in the range of $0.5 \leqslant n \leqslant 1.5$, the relation of $\mu_{0}$ from two models can be written as $\mu_{\rm 0,s\acute{e}rsic} - \mu_{\rm 0,exp} = -1.34(n-1)$.
The LSBGs selected by disk+bulge models (LSBG\_2comps) are more massive than LSBGs selected by single-component models (LSBG\_1comp), and also show a larger disk component. Though the bulges in the majority of our LSBG\_2comps are not prominent, more than 60\% of our LSBG\_2comps will not be selected if we adopt a single-component model only.
%About 1/4 of the LSBGs selected by exp+ser model hold a classical bulge and the rest hold a pseudo bulge. LSBGs with pseudo bulge tend to have smaller B/T and effective radius ratio of bulge and disk components than LSBGs with classical bulge.
We also identified 31 giant low surface brightness galaxies (gLSBGs) from LSBG\_2comps. They are located at the same region in the color-magnitude diagram as other gLSBGs. After we compared different criteria of gLSBGs selection, we find that for gas-rich LSBGs, $M_{\star} > 10^{10}M_{\odot}$ is the best to distinguish between gLSBGs and normal LSBGs with bulge.
}

\authorrunning{B.-Q. Zhang et al.}            %author_head in even pages
\titlerunning{Low Surface Brightness Galaxies selected by different model fitting}  % title_head in odd pages
\maketitle

%________________________________________________ sections below
% 
\section{Introduction} \label{introduction}
Low surface brightness galaxies (LSBGs) are galaxies whose central surface brightness of the disk component is at least one magnitude fainter than the sky brightness \citep{1997ARAA...35..267I}. Studies show that the LSBGs are under-evolved because they generally have sparse H$\upalpha$ emission \citep{1997AJ....114.1858P, 2014ApJ...793...40H}, low star formation rate (SFR), low SFR surface density \citep{1993AJ....106..548V, 2011ApJ...728...74G, 2018ApJS..235...18L, 2019ApJS..242...11L}, and low metallicity \citep{1994AJ....107..530M, 2010MNRAS.409..213L}. They are rich in neutral hydrogen ($\rm H\Rmnum{1}$) gas \citep{1996MNRAS.283...18D, 2001AJ....122.2318B, 2004AJ....128.2080O, 2015AJ....149..199D}, but have little CO molecules \citep{2003ApJ...588..230O, 2018MNRAS.476.4488H}, indicating they have low efficiency in converting H\Rmnum{1} gas to $\rm H_{2}$ molecules \citep{2017AJ....154..116C}.

LSBGs are mostly late-type disk-dominated galaxies \citep{1994AJ....107..530M, 1995MNRAS.274..235D}. \cite{1997AJ....113.1212O} show around 80\% of the galaxies in their LSBGs sample are well-fitted by an exponential profile. So it is appropriate to fit the LSBGs with a single disk model, many authors have done so \citep{1997AJ....113.1212O, 2015AJ....149..199D}. However, \cite{1995AJ....109.2019M} claim that a small but significant subset of LSBGs has $\rm B/D \approx 1$ (bulge-to-disk ratio), \cite{2018MNRAS.478.4657P} show that about 40\% of the galaxies in their LSBGs sample are with bulges. This kind of LSBGs that have bulges at their center will be lost if only a single disk model is adopted. Then two approaches arose, one is to process a bulge+disk decomposition to get the central surface brightness of disk component \citep{2008MNRAS.387.1099P, 2018MNRAS.478.4657P}, another is to use the average surface brightness within effective radius ($\rm \bar{\mu}_{e}$) instead of central surface brightness ($\mu_{0}$) to reduce the effect of luminosity concentration in the galaxy center \citep{2018ApJ...857..104G, 2019MNRAS.485..796M, 2021ApJS..252...18T}. The second approach cannot get detailed information of the disk component, motivating us to search for LSBGs with bulges by using the bulge+disk decomposition method.

In another aspect, the observation technology improved and the wide-field surveys developed a lot in the last two decades, e.g., the Sloan Digital Sky Survey (SDSS; \citealt{2000AJ....120.1579Y}), the Dark Energy Camera Legacy Survey (DECaLS; \citealt{2019AJ....157..168D}), the Arecibo Legacy Fast ALFA survey (ALFALFA; \citealt{2005AJ....130.2598G}). People can dig deeper and wider in the sky, also bringing more opportunities for us to search for LSBGs. For example, 12282 LSBGs were selected from SDSS data release 4 (DR4) to study their stellar population \citep{2008MNRAS.391..986Z} and metallicities \citep{2010MNRAS.409..213L}; \citet{2011ApJ...728...74G} also selected 9421 LSBGs from SDSS DR4 to investigate their spatial distribution. The combination of optical images and H\Rmnum{1} spectra provides us with one of the best laboratories for studying gas-rich LSBGs. \cite{2011AJ....142..170H} provide a cross-reference catalog, which contains 12468 galaxies, of $\upalpha$.40 (40\% sky area of the full ALFALFA) and SDSS Data Release 7 (DR7; \citealt{2009ApJS..182..543A}), and \cite{2015AJ....149..199D} (hereafter \citetalias{2015AJ....149..199D}) developed a pipeline to reestimate the sky background for 12423 galaxies belonging to the PhotoPrimary catalog in the $\upalpha$.40 SDSS DR7 sample to avoid the sky background overestimation by the photometric pipeline of the SDSS. Based on the background resubtracted images, \citetalias{2015AJ....149..199D} selected a none-edge-on LSBG sample of 1129 galaxies by using a single exponential model fitting, this paper focuses on selecting LSBGs from different models and searching for LSBGs with bulges by applying disk+bulge model.

The rest of the paper is arranged as follows. In Section \ref{Sample and fitting}, we describe our parent sample, model fitting, and the method of central surface brightness ($\mu_{0}$) calculation. Section \ref{results} shows our LSBGs samples and their statistical properties. In Section \ref{gLSBGs}, we discuss the gLSBGs selection. Finally, we summarize this paper in Section \ref{summary}.

\section{Sample and fitting} \label{Sample and fitting}
\subsection{Parent sample} \label{Parent sample}
Many researchers, including the ones in the SDSS team, noticed that the sky background in the SDSS imaging pipeline is overestimated \citep{2006ApJS..162...38A, 2008ApJS..175..297A, 2007ApJ...662..808L, 2008MNRAS.385...23L, 2009MNRAS.394.1978H, 2013ApJ...773...37H}, leading to the galaxy brightness being underestimated. Because the central surface brightness of LSBGs is fainter than the sky background, the outer parts of LSBGs are even fainter. The inaccurate background will influence the photometry, model fitting, and LSBG selection. \citetalias{2015AJ....149..199D} carefully estimated the sky background of SDSS images in both g and r bands for 12,423 galaxies in the $\upalpha$.40 SDSS DR7 sample \citep{2011AJ....142..170H}. After their background subtraction, the count distributions for the whole image and the local vicinity, which defined as the region between the two square boxes sized 250 $\times$ 250 pixels and 500 $\times$ 500 pixels from the galaxy center, are both well fitted by a Gaussian profile with mean values very close to 0 ADU (see Figure 3 in \citetalias{2015AJ....149..199D}). We take the 12,423 galaxies as our parent sample in selecting LSBGs, taking advantage of their better analysis on the sky subtraction.

We took photometry for all galaxies in our parent sample by using SExtractor software \citep{1996A&AS..117..393B}. A flexible elliptical aperture, the Kron aperture defined by \citet{1980ApJS...43..305K}, was used to get magnitude (MAG\_AUTO). The apertures of g and r bands are the same which were defined by the r band image. The parameters from SExtractor, including magnitude, effective radius, axis ratio, and position angle, are used as initial guess for model fitting.

\subsection{Model fitting} \label{Galfit fitting}
Galfit is a two-dimensional fitting algorithm to extract structural components of galaxies \citep{2002AJ....124..266P, 2010AJ....139.2097P}. It provides some of the most commonly used radial profiles in astronomy literature, e.g., the exponential, s\'{e}rsic, deVaucular, gaussian, moffat, and psf profiles. The exponential and deVaucular profiles are special cases of the s\'{e}rsic function when $n = 1$ and $n = 4$, respectively. Users can adopt a single profile or a combination of a number of profiles, and set initial values for input parameters of each profile. During the fitting, the reduced $\chi^{2}$ is minimized, and the minimization engine is based on the Levenberg-Marquardt downhill gradient algorithm.

We carried out two kinds of single-component fitting and two kinds of two-component (disk+bulge) fitting: the single exponential profile, the single s\'{e}rsic profile, combination of the exponential and deVaucular profile (hereafter, exp+deV), and combination of the exponential and s\'{e}rsic profile (hereafter, exp+ser). Since most galaxies in our parent sample can be fully shown in 501 $\times$ 501 pixels, all the fittings are performed in this region with the target located at the center to improve Galfit fitting efficiency. During fitting, psf image and mask image were used to get better results. The psf images were derived from SDSS website (http://das.sdss.org/imaging) to describe the local psf profile surrounding the target, and the mask images were produced on the basis of the segmentation image from SExtractor. We masked out other objects, only the background and target regions are fitted. 

We also set limitations on variable range of parameters to avoid unphysical fittings. For all the models and subcomponents, the variable range of the galaxy center is $\pm 5$ pixels in both horizontal and vertical directions. Based on the assumption that half of the diagonal size of our images($\sim$ 354 pixel) being exactly three times the $\rm R_{e}$ of the galaxy, we set the variable range of the effective radius ($\rm R_{e}$) as $1 \sim 118$ pixels, that is $1\leq {\rm R_{e}}\leq118$ for s\'{e}rsic and deVaucular profile and $1\leq 1.678\times {\rm R_{s}}\leq 118$ for the exponential profile, where $\rm R_{s}$ is scale length. Here we use $\rm R_{s}$ instead of $\rm R_{e}$ for exponential profile is because the exponential profile provided by Galfit (Equation \ref{exp_profile}) uses $\rm R_{s}$, and there is a relation between $\rm R_{e}$ and $\rm R_{s}$ for exponential profile, $\rm R_{e}\ =\ 1.678\times R_{s}$ \citep[Eq. (7)]{2010AJ....139.2097P}. For the s\'{e}rsic model, the variable range of $n$ index is $0.5 \sim 8.0$, which is the same range as \citet{2011ApJS..196...11S}. When $n < 0.5$, the luminosity density of the s\'{e}rsic profile has a depression in its center, which is often unphysical \citep[see][Fig. 6]{2001MNRAS.321..269T}. For the exp+ser model, the variable range of the $n$ index of the s\'{e}rsic component is $1.0 \sim 8.0$, here we force the bulge component (s\'{e}rsic) to be more concentrated than the disk component (exponential) by applying a larger $n$ value of the s\'{e}rsic profile than that of the exponential profile.
%we expect the s\'{e}rsic profile to stand for the bulge component, so

%We selected galaxies with n > 1.5 in the single sérsic fitting to perform two-component fitting, this is because the galaxies with 0.5 < n < 1.5 are disk-dominated galaxy, which can be well described by a single disk model. On the contrary, the application of a two-component model by force would lead Galfit to converge to no results.

Note that all the galaxies have been fitted with the exponential and the s\'{e}rsic profile, then 5233 galaxies (42.12\% of parent sample) were selected to run the disk+bulge model fitting by eliminating 7190 galaxies whose s\'{e}rsic $n$ index $0.5 \leqslant n \leqslant 1.5$ for both g and r bands. This is because, for disk-dominated galaxies, the application of a disk+bulge model by force would lead Galfit to converge to no results.

After fitting, we took the following steps to filter the fitting results:
\begin{enumerate}
\item Eliminate numerical unreasonable fittings. When numerical convergence issues happen during fitting, the entire solution is not reliable, we will remove the galaxy from further analysis. The galaxy will also be removed if the output parameters reach the boundaries we set. In this step, for s\'{e}rsic and exp+ser models, most of the removed galaxies are because their fitted $n$ index are at the boundary values.
\item Eliminate fitting results whose structure parameters, such as the axis ratio (b/a) and position angle (PA), are not consistent between the g and r bands. For b/a and PA, we calculate the difference for all the galaxies and execute a 3$\upsigma$ clipping. 
\item For disk+bulge fitting, galaxies will be removed if their fitted effective radius of the disk component is smaller than those of the bulge component.
\end{enumerate}

After these steps, we constructed four ``well-fitted" samples which contain 11697, 10582, 1981, and 1066 galaxies for exponential, s\'{e}rsic, exp+deV, and exp+ser model fitting, respectively. The galaxies in the ``well-fitted" samples have good fitting results in both g and r bands, and will be used in the following analysis. The numbers of galaxy after each step are listed in Table \ref{quantities}. 

\begin{table}[h]
\centering
\caption{Sample Numbers after each Step.} \label{quantities}
\begin{threeparttable}
\begin{tabular}{ccccccccc}
\hline
 & \multicolumn{2}{c}{exponential} & \multicolumn{2}{c}{s\'{e}rsic} & \multicolumn{2}{c}{exp+deV} & \multicolumn{2}{c}{exp+ser} \\ \cline{2-9}
 & g & r & g & r & g & r & g & r \\
\hline
feed in Galfit\tnote{1} & 12423 & 12423 & 12423 & 12423 & 5233  & 5233  & 5233  & 5233  \\
have Galfit output & 12218 & 12218 & 12217 & 12217 & 4560  & 4665  & 4461  & 4680  \\ numerical reasonable\tnote{2} & 12058 & 12088 & 11096 & 11353 & 3592  & 3800  & 2045  & 2379  \\ \hline
good structure parameters\tnote{3}  & \multicolumn{2}{c}{11697}   & \multicolumn{2}{c}{10582}   & \multicolumn{2}{c}{3031} & \multicolumn{2}{c}{1495}  \\
$\rm R_{e,disk} > R_{e,bulge}$ & \multicolumn{2}{c}{---} & \multicolumn{2}{c}{---} & \multicolumn{2}{c}{1981} & \multicolumn{2}{c}{1066}  \\ \hline
well-fitted sample & \multicolumn{2}{c}{11697}   & \multicolumn{2}{c}{10582}   & \multicolumn{2}{c}{1981} & \multicolumn{2}{c}{1066}  \\ 
LSBG sample (initial) & \multicolumn{2}{c}{1151}    & \multicolumn{2}{c}{1081}    & \multicolumn{2}{c}{286}  & \multicolumn{2}{c}{175}   \\ 
LSBG sample (after visual inspection) & \multicolumn{2}{c}{1105}    & \multicolumn{2}{c}{1038}    & \multicolumn{2}{c}{207}  & \multicolumn{2}{c}{132}  \\ 
LSBG sample ($n_{\rm b}>1.5$)\tnote{4} & \multicolumn{2}{c}{---} & \multicolumn{2}{c}{---} & \multicolumn{2}{c}{---} & \multicolumn{2}{c}{75} \\ \hline
\end{tabular}
\begin{tablenotes}
\item[1] The number of galaxies that we feed in Galfit for exp+deV and exp+ser fitting excludes the galaxies that satisfy $0.5 \leqslant n \leqslant 1.5$ for both g and r bands in s\'{e}rsic fitting.
\item[2] The numerical reasonable fittings are fittings whose $\rm R_{e}$, $\rm R_{s}$, and $n$ index do not reach the boundaries.
\item[3] The structure parameters mean the axis ratio (b/a) and position angle (PA). We calculate the difference between g and r bands and execute a 3$\upsigma$ clipping, for b/a and PA one by one. The fittings with good structure parameters mean their b/a and PA are consistent between the g and r bands.
\item[4] The $n_{\rm b}$ represents the $n$ index of bulge component.
\end{tablenotes}
\end{threeparttable}
\end{table}

%g and r band match & \multicolumn{2}{c}{12017}   & \multicolumn{2}{c}{10884}   & \multicolumn{2}{c}{3209} & \multicolumn{2}{c}{1575}  \\

\subsection{The central surface brightness} \label{Calculate mu0}
In the following we will illustrate the procedures how we calculate $\mu_{0}$ for the exponential profile and the s\'{e}rsic profile step by step.

The expression of the exponential profile is:
\begin{equation}
\rm
f(r) = f_{0}exp(-\frac{r}{r_{s}})
\label{exp_profile}
\end{equation}

The flux integrated out to $\rm r = \infty$ is:
\begin{equation}
\rm
F_{total} = 2\pi r_{s}^{2}f_{0}q
\end{equation}
where q is the axis ratio.

So the central surface brightness can be expressed as:
\begin{equation}
\rm
f_{0} = \frac{F_{total}}{2\pi r_{s}^{2}q}
\end{equation}

Then, apply cosmological dimming correction, also convert the unit of $\rm f_{0}$ to $\rm mag\ arcsec^{-2}$, the central surface brightness is:
\begin{align}
\mu_{0} &= {\rm -2.5log_{10}(f_{0}) - 10log_{10}}(1 + z) \nonumber \\
        &= m + {\rm 2.5log_{10}(2\pi r_{s}^{2}q) - 10log_{10}}(1 + z)
\end{align}
where $m$ is the model magnitude, $\rm r_{s}$ is the disk scale length, and $z$ is from $\upalpha$.40 catalog.

For s\'{e}rsic model, the function is:
\begin{equation}
{\rm f(r) = f_{e} exp[-\kappa((\frac{r}{r_{e}})}^{\frac{1}{n}} - 1)]
\label{sersic_profile}
\end{equation}
%\textcolor{blue}{\sout{The exponential and deVaucular profiles are special cases of the s\'{e}rsic function when n = 1 and n = 4, respectively.}}

The flux integrated out to $\rm r = \infty$ is:
\begin{equation}
{\rm F_{total} = 2\pi r_{e}^{2}f_{e}}e^{\kappa}n\kappa^{-2n}\Gamma(2n){\rm q}
\end{equation}

So the surface brightness at the effective radius($\rm f_{e}$) and galaxy center($\rm f_{0}$) is:
\begin{align}
{\rm f_{e}} &= \frac{\rm F_{total}}{{\rm 2\pi r_{e}^{2}}e^{\kappa}n\kappa^{-2n}\Gamma(2n){\rm q}}  \\
{\rm f_{0}} &= {\rm f_{e} exp(\kappa)} = \frac{\rm F_{total}}{{\rm 2\pi r_{e}^{2}}n\kappa^{-2n}\Gamma(2n){\rm q}} 
\end{align}
where $\rm r_{e}$ is the effective radius, $n$ is the s\'{e}rsic index, $\kappa$ is related to $n$ index $\kappa = 1.9992n - 0.3271$ \citep{1989woga.conf..208C}, $\Gamma$ function is $\Gamma(2n) = (2n - 1)!$.

Then, apply cosmological dimming correction, also convert the unit of $\rm f_{0}$ to $\rm mag\ arcsec^{-2}$, the central surface brightness is:
\begin{align}
\mu_{0} &= {\rm -2.5log_{10}(f_{0}) - 10log_{10}}(1 + z) \nonumber \\
        &= m + {\rm 2.5log_{10}(2\pi r_{e}^{2}}n\kappa^{-2n}\Gamma(2n){\rm q}) - {\rm 10log_{10}}(1 + z) 
\end{align}

Finally, the B band central surface brightness can be transferred by an empirical equation \citep{2002AJ....123.2121S}:
\begin{equation}
\rm 
\mu_{0}(B) = \mu_{0}(g) + 0.47(\mu_{0}(g) - \mu_{0}(r)) + 0.17  \label{mu0B}
\end{equation}

\begin{figure}[h]
    \centering
    \includegraphics[width=\textwidth]{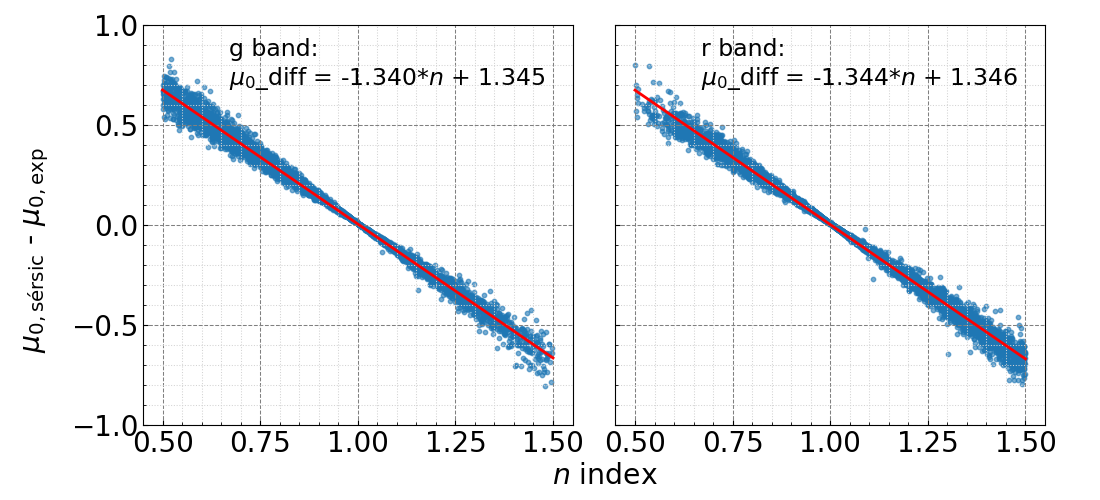}
    \caption{The relation between the difference of central surface brightness from exponential and s\'{e}rsic fitting and the s\'{e}rsic $n$ index. The red solid lines are the linear fitting of the points. The fitting functions are written on the top of each panel.}
    \label{transfer_mu0}
\end{figure}

Figure \ref{transfer_mu0} shows the difference of $\mu_{0}$ between exponential and s\'{e}rsic fitting varies with s\'{e}rsic $n$ index for galaxies which have $0.5 \leqslant n \leqslant 1.5$ in both g and r bands. The red solid lines are the linear fitting of the points, which can be described as follows:
\begin{align}
{\rm g\ band: \mu_{0,s\acute{e}rsic} - \mu_{0,exp}} &= -1.340n + 1.345   \\
{\rm r\ band: \mu_{0,s\acute{e}rsic} - \mu_{0,exp}} &= -1.344n + 1.346   
\end{align}
which also be written on the top of each panel. The expression can be simplified to 
\begin{equation}
{\rm \mu_{0,s\acute{e}rsic} - \mu_{0,exp}} = -1.34(n-1)
\end{equation}
The relationship between $\mu_{0}$ difference and $n$ index distributes along a biconical shape, which is best when $n = 1$, and becomes worse with the increase or decrease of $n$ index. When $n > 1$, the $\mu_{0}$ obtained from the s\'{e}rsic model is brighter than that from the exponential model, and when $n < 1$, the $\mu_{0}$ obtained from the s\'{e}rsic model is dimmer than that from the exponential model. The two models yield a $\mu_{0}$ difference of up to approximately 0.7 $\rm mag\ arcsec^{-2}$.

\section{LSBG samples} \label{results}
\subsection{LSBG selection} \label{lsbg selection}
There has been no consensus in the threshold values of $\mu_{0}$ to define LSBGs, people apply different cut in literature ($\rm \mu_{0,disk}(B) = 22.0\ mag\ arcsec^{-2}$ \citep{1995AJ....110..573M}, $\rm 22.5\ mag\ arcsec^{-2}$ (\citealt{1996IAUS..171...97M, 2009A&A...504..807R}; \citetalias{2015AJ....149..199D}), $\rm 23.0\ mag\ arcsec^{-2}$ \citep{1997ARAA...35..267I}; $\rm \mu_{0,disk}(V) = 21.2\ mag\ arcsec^{-2}$ \citep{2001AJ....122..714B}; $\rm \mu_{0,disk}(R) = 20.8\ mag\ arcsec^{-2}$ \citep{1996ApJS..103..363C, 2001AJ....122..714B}; $\rm \mu_{0,disk}(r) = 21.0\ mag\ arcsec^{-2}$ \citep{2018MNRAS.478.4657P}). In this paper, we adopt a threshold of $\rm \mu_{0,disk}(B) \geqslant 22.5\ mag\ arcsec^{-2}$ and apply $\rm b/a \geqslant 0.3$ to avoid internal extinction effect \citep{2020ApJS..248...33H}.

With this selection criterion, and the visual inspection afterwards, we selected four LSBG samples from the ``well-fitted sample" of each model: LSBG\_exp, LSBG\_s\'{e}rsic, LSBG\_exp+deV, and LSBG\_exp+ser, which contain 1105 galaxies (9.45\% of ``exponential'' well-fitted sample), 1038 galaxies (9.81\% of ``s\'{e}rsic'' well-fitted sample), 207 galaxies (10.45\% of ``exp+deV'' well-fitted sample), and 132 galaxies (12.38\% of ``exp+ser'' well-fitted sample). During the visual inspection, fittings were deleted if the model was influenced obviously by the neighbors or the galaxy is peculiar in morphology. Considering that our original intention was to look for LSBGs with bulge, we set a limitation on the $n$ index of bulge component of ``exp+ser" model $n_{\rm b} > 1.5$ to exclude some ``disk+disk" fitting results, and there are 75 LSBGs satisfy this criterion in both g and r bands. The galaxy numbers of initial LSBG samples and final LSBG samples are listed in Table \ref{quantities}. The table containing all the parameters of our LSBGs (DOI: https://doi.org/10.57760/sciencedb.13130) is available in its entirety in a machine-readable form in the Science Data Bank database, the column descriptions are listed in Table \ref{table_LSBG}.

\begin{figure}[h]
    \centering
    \includegraphics[width=\textwidth]{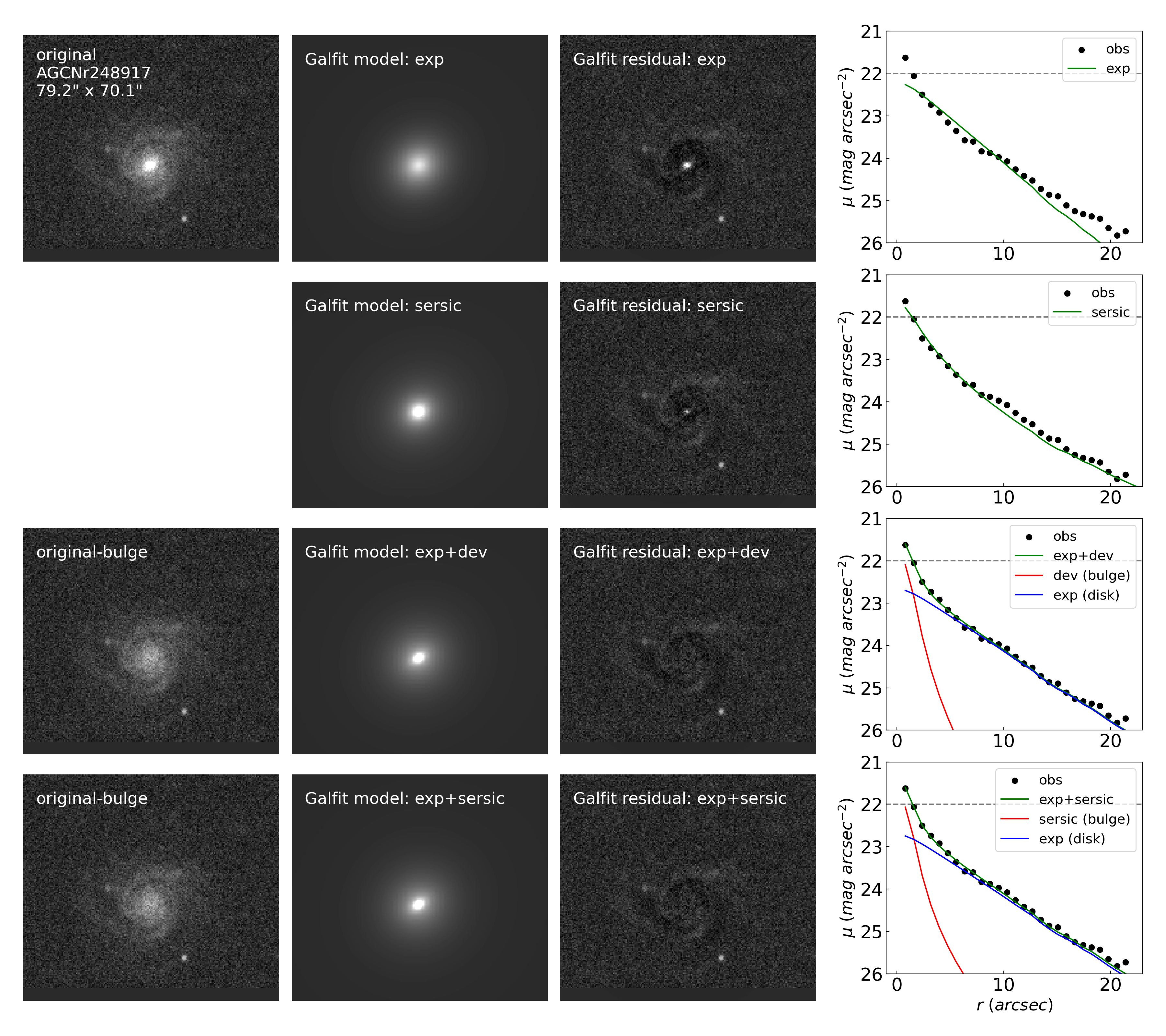}
    \caption{An example of Galfit fitting results of four models, exponential, s\'{e}rsic, exp+deV, and exp+ser, from top to bottom. This galaxy is selected as LSBG by ``exp+deV" and ``exp+ser" models. The upper left corner shows the observed image, the second column shows the model images, the third column shows the residual images (observed image minus model image), and the fourth column is the radial distribution of surface brightness obtained from a series of elliptical annulus on the observed image and models, with a step of 2 pixels. The bottom two rows of the first column show the pure disk image (observed image minus bulge component).}
    \label{compare_gal_fitting_1comp_2comps}
\end{figure}

For the LSBGs selected from two-component fitting, the two-component fitting is usually better than the one-component fitting. Figure \ref{compare_gal_fitting_1comp_2comps} shows an example of Galfit fitting results (AGCNr 248917) of our four models, exponential, s\'{e}rsic, exp+deV, and exp+ser, from top to bottom. This galaxy is selected as LSBG by ``exp+deV" and ``exp+ser" models. The first image in the upper left panel shows the observed image, the second column shows the model images, the third column shows the residual images (observed image minus model image), and the fourth column is the radial distribution of surface brightness obtained from a series of elliptical annulus on the observed image and models, with a step of 2 pixels. The pure disk image is also displayed in the bottom two rows of the first column by using the observed image minus the bulge component. From the residual image of exponential and s\'{e}rsic fitting, it is obvious that there is a small bright spot in the galaxy center, and this small bright spot can be fitted well by exp+deV and exp+ser models. 

%figure: 2comps is better than 1comp
%Figure \ref{} shows an example of ``disk+bulge" galaxy and ``disk+disk" galaxy. 

\begin{table}[!h]
\centering
\caption{Column descriptions of our LSBG table.} \label{table_LSBG}
%\resizebox{\textwidth}{!}{
\begin{threeparttable}
\begin{tabular}{ccc}
\hline
Column name & Description & Unit \\
\hline
note                  & Model adopt                                                        &                        \\
AGCNr                 & ID in $\upalpha$.40 catalog                                        &                        \\
ra                    & Right ascension (J2000.0)                                          & degree                 \\
dec                   & decl. (J2000.0)                                                    & degree                 \\
Vhelio                & Heliocentric velocity of the HI profile midpoint from $\upalpha$.40 catalog & km/s          \\
Dist                  & Distance from $\upalpha$.40 catalog                                & Mpc                    \\
logMHI                & HI mass from $\upalpha$.40 catalog                                 & logarithmic solar mass \\
W50                   & Observed velocity width at 50\% of peak on either side from $\upalpha$.40 catalog & km/s    \\
kpc\_per\_arcsec      & image scale factor                                                 & kpc/arcsec             \\
kpc\_per\_pix         & image scale factor                                                 & kpc/pixel              \\
magauto\_all\_band    & Magnitude for the whole galaxy from SExtractor                     & mag                    \\
magautoerr\_all\_band & Error of magnitude for the whole galaxy from SExtractor            & mag                    \\
MAG\_obs\_band        & Absolute magnitude for the whole galaxy                            & mag                    \\
M\_star\_obs\_band    & Stellar mass for the whole galaxy                                  & logarithmic solar mass \\
mag\_disk\_band       & Magnitude of disk component from model fitting                     & mag                    \\
mag\_disk\_err\_band  & Error of magnitude of disk component from model fitting            & mag                    \\
rs\_disk\_band        & Disk-scale length of disk component from model fitting             & pixel                  \\
rs\_disk\_err\_band   & Error of disk-scale length of disk component from model fitting    & pixel                  \\
re\_disk\_band        & Effective radius of disk component from model fitting              & pixel                  \\
re\_disk\_err\_band   & Error of effective radius of disk component from model fitting     & pixel                  \\
n\_disk\_band         & s\'{e}rsic $n$ index of disk component from model fitting          &                        \\
n\_disk\_err\_band    & Error of s\'{e}rsic $n$ index of disk component from model fitting &                        \\
ar\_disk\_band        & Axis ratio of disk component from model fitting                    &                        \\
ar\_disk\_err\_band   & Error of axis ratio of disk component from model fitting           &                        \\
pa\_disk\_band        & Position angle of disk component from model fitting                & degree: Up=0   Left=90 \\
pa\_disk\_err\_band   & Error of position angle of disk component from model fitting       & degree: Up=0   Left=90 \\
mag\_bulge\_band      & Magnitude of bulge component from model fitting                    & mag                    \\
mag\_bulge\_err\_band & Error of magnitude of bulge component from model fitting           & mag                    \\
re\_bulge\_band       & Effective radius of bulge component from model fitting             & pixel                  \\
re\_bulge\_err\_band  & Error of effective radius of bulge component from model fitting    & pixel                  \\
n\_bulge\_band        & s\'{e}rsic $n$ index of bulge component from model fitting         &                        \\
n\_bulge\_err\_band   & Error of s\'{e}rsic $n$ index of bulge component from model fitting &                       \\
ar\_bulge\_band       & Axis ratio of bulge component from model fitting                   &                        \\
ar\_bulge\_err\_band  & Error of axis ratio of bulge component from model fitting          & degree: Up=0   Left=90 \\
pa\_bulge\_band       & Position angle of bulge component from model fitting               & degree: Up=0   Left=90 \\
pa\_bulge\_err\_band  & Error of position angle of bulge component from model fitting      &                        \\
mu0\_disk\_band       & Central surface brightness of disk component                       & $\rm mag\ arcsec^{-2}$ \\
mu0\_disk\_err\_band  & Error of central surface brightness of disk component              & $\rm mag\ arcsec^{-2}$ \\
B/T\_band             & Bulge-to-total ratio from model fitting                            &                        \\
\hline
\end{tabular}
\begin{tablenotes}
\footnotesize
\item \normalsize This is column descriptions of our LSBG table. The related table (DOI: https://doi.org/10.57760/sciencedb.13130) is available in its entirety in a machine-readable form in the Science Data Bank database.
\end{tablenotes}
\end{threeparttable}
%}
\end{table}

\subsection{Statistical properties of LSBG samples} \label{Statistic properties of LSBGs}
\begin{figure}[h]
    \centering
    \includegraphics[width=\textwidth]{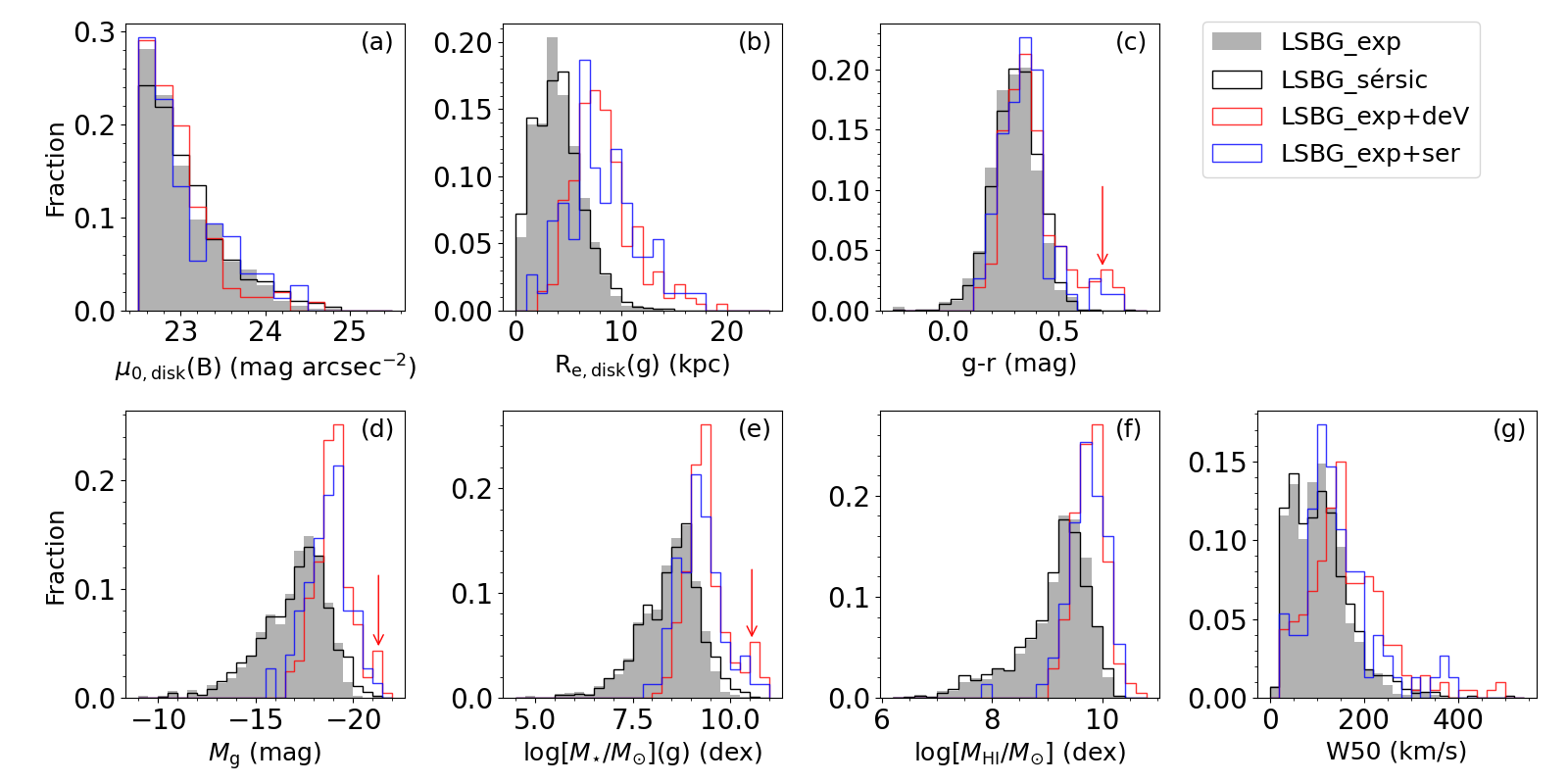}
    \caption{The statistical properties of our four LSBG samples. The B band central surface brightness ($\rm \mu_{0,disk}(B)$) and g band effective radius ($\rm R_{e,disk}(g)$) of the disk component are from model fitting; the g-r color, g band absolute magnitude ($M_{\rm g}$), and g band stellar mass (log$(M_{\star}/M_{\odot})$(g)) are from photometry; the H\Rmnum{1} gas mass and W50 are from the $\upalpha$.40 catalog \citep{2011AJ....142..170H}. The red arrows in panels (c), (d), and (e) point out another peak of distributions of LSBG\_exp+deV and LSBG\_exp+ser samples.}
    \label{LSBG_all}
\end{figure}

Figure \ref{LSBG_all} shows the statistical properties of our four LSBG samples: panels (a) and (b) show the distribution of the B band central surface brightness ($\rm \mu_{0,disk}(B)$) and g band effective radius ($\rm R_{e,disk}(g)$) of the disk component; panels (c)$\sim$(g) show the distribution of g-r color, g band absolute magnitude ($M_{\rm g}$), g band stellar mass (log$(M_{\star}/M_{\odot})$(g)), H\Rmnum{1} gas mass (log$(M_{\rm H\Rmnum{1}}/M_{\odot})$), and H\Rmnum{1} line width (W50) for the whole galaxy. The H\Rmnum{1} gas mass and W50 are from the $\upalpha$.40 catalog \citep{2011AJ....142..170H}. The stellar mass was calculated by using the prescription in \cite{2020AJ....159..138D}. In that, we compute mass-to-light ratio by using the g-r color: $\rm log_{10}(M/L)_{\lambda} = a_{\lambda} + (b_{\lambda} \times (g-r))$, where $\rm a_{g} = -0.857$, $\rm b_{g} = 1.558$, $\rm a_{r} = -0.7$, and $\rm b_{r} = 1.252$. The absolute magnitude was calculated by using the distance from the $\upalpha$.40 catalog by adopting $\rm H_{0} = 70\ km\ s^{-1}\ Mpc^{-1}$ \citep{2011AJ....142..170H}, and the absolute magnitude of the sun is $M_{\rm g, \odot} = 5.11$ and $M_{\rm r, \odot} = 4.65$ \citep{2018ApJS..236...47W}. We list the median values of these parameters of our four LSBG samples in Table \ref{median_values}. Table \ref{in common} lists the number statistics of the overlapping galaxies that are both selected as LSBGs in any of the two LSBG samples.

\begin{table}[h]
\centering
\caption{The Median Values of Parameters of our Four LSBG Samples.} \label{median_values}
\begin{tabular}{cccccccc}
\hline
& $\rm \mu_{0,disk}(B)$ & $\rm R_{e,disk}(g)$ & g-r & $M_{\rm g}$ & log$(M_{\star}/M_{\odot})$(g) & log$(M_{\rm H\Rmnum{1}}/M_{\odot})$ & W50 \\
& ($\rm mag\ arcsec^{-2}$) & (kpc) & (mag) & (mag) & ($M_{\odot}$) & ($M_{\odot}$) & (km/s) \\
\hline
LSBG\_exp        & 22.888 & 3.843 & 0.298 & $-17.257$ & 8.539 & 9.28 & 100.0 \\
LSBG\_s\'{e}rsic & 22.936 & 3.840 & 0.311 & $-17.294$ & 8.561 & 9.25 & 100.0 \\
LSBG\_exp+deV    & 22.882 & 7.902 & 0.354 & $-18.965$ & 9.305 & 9.76 & 149.0 \\
LSBG\_exp+ser    & 22.897 & 7.665 & 0.343 & $-18.774$ & 9.195 & 9.72 & 127.0 \\
%LSBG\_1comp      & 22.829 & 3.909 & 0.308 & -17.421 & 8.615 & 9.3  & 103.0 \\
%LSBG\_2comps     & 22.882 & 7.815 & 0.346 & -18.913 & 9.275 & 9.74 & 146.0 \\
\hline
\end{tabular}
\end{table}

\begin{table}[h]
\centering
\caption{The Numbers of Galaxies Duplicated in any Two LSBG Samples.} \label{in common}
\begin{tabular}{c|ccccc}
\hline
 & LSBG\_exp & LSBG\_s\'{e}rsic & LSBG\_exp+deV & LSBG\_exp+ser & LSBG\_exp+ser ($n>2$) \\
\hline
LSBG\_exp        & 1105      & 756              & 76            & 28            & 13  \\
LSBG\_s\'{e}rsic & 756       & 1038             & 15            & 4             & 0   \\
LSBG\_exp+deV    & 76        & 15               & 207           & 35            & 22  \\
LSBG\_exp+ser    & 28        & 4                & 35            & 75            & 37  \\
LSBG\_exp+ser ($n>2$) & 13   & 0                & 22            & 37            & 37  \\
\hline
\end{tabular}
\end{table}

% the same and the difference between LSBG_exp and LSBG_sersic
\subsubsection{Comparison of LSBG\_exp and LSBG\_s\'{e}rsic}
\begin{figure}[h]
    \centering
    \includegraphics[width=8cm]{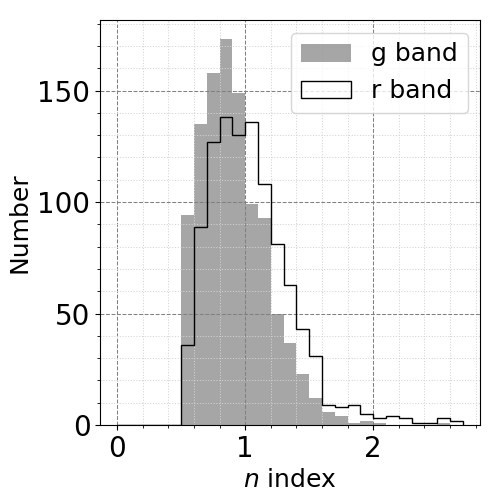}
    \caption{The distribution of $n$ index for galaxies in LSBG\_s\'{e}rsic. Around 97.40\% of them have $0.5 < n_{\rm g} < 1.5$ and 91.62\% have $0.5 < n_{\rm r} < 1.5$.}
    \label{hist_n}
\end{figure}

It is obvious to see that all the parameters of LSBG\_exp (the gray filled histograms) and LSBG\_s\'{e}rsic (the black steps) have very similar distributions, and the median values of these parameters are close. Figure \ref{hist_n} shows the distribution of $n$ index of galaxies in LSBG\_s\'{e}rsic, 97.40\% (1011/1038) of the galaxies have $0.5 < n < 1.5$ for g band and 91.62\% (951/1038) for r band, indicating our LSBGs selected by s\'{e}rsic fitting are in principle the same type of galaxies as those selected by exponential fitting. These two samples have 756 galaxies (68.42\% of LSBG\_exp and 72.83\% of LSBG\_s\'{e}rsic) in common, the rest of the discrepancy is due to the difference in obtaining $\mu_{0}$ between the two models as we mentioned in Section \ref{Calculate mu0}. That is, the remaining galaxies have $\mu_{0}$ brighter than 22.5 $\rm mag\ arcsec^{-2}$ in one fitting result and dimmer than 22.5 $\rm mag\ arcsec^{-2}$ in the other fitting result.

% the same and the difference between LSBG_exp+dev and LSBG_exp+ser
\subsubsection{Comparison of LSBG\_exp+deV and LSBG\_exp+ser}
The distributions of all the parameters of the galaxies in LSBG\_exp+deV (the red steps) and LSBG\_exp+ser (the blue steps) are also similar, but the median values of these parameters of LSBG\_exp+deV are slightly larger than those of LSBG\_exp+ser.

\cite{2008AJ....136..773F} found that the s\'{e}rsic index of bulge can be used to distinguish between classical and pseudo bulges: classical bulges have $n_{\rm b} \gtrsim 2$ and pseudo bulges have $n_{\rm b} \lesssim 2$ in the V band with almost no overlap. Considering that the g band has more similar wavelength coverage to the V band than the r band, we divide the bulges in the LSBG\_exp+ser sample into classical and pseudo bulges according to their $n$ index of bulge component ($n_{\rm b}$) in g band. Of these 75 galaxies, 50 galaxies are classified as galaxies with classical bulges.

Among the 50 LSBGs that hold classical bulges, 24 of them are selected as LSBGs by ``exp+deV'' fitting. In the rest 26 galaxies in LSBG\_exp+ser, 25 of them are not in the ``exp+deV'' well-fitted sample, only one galaxy (AGCNr 102234) is in but not selected as LSBG because it does not satisfy the $\rm b/a \geqslant 0.3$ criterion. The reason why one model can get the fitting result but the other model cannot is beyond the scope of this paper. Only consider the galaxies that both models can get good fitting results, almost all the LSBGs with classical bulges selected by the ``exp+ser'' model can be selected by the ``exp+deV'' model.
%the selected LSBGs are consistent between ``exp+deV" and ``exp+ser" models. A
%\textcolor{blue}{\sout{The contours in Figure \ref{CMR_gLSBG} also show that LSBG\_exp+deV contains some red and bright (or massive) LSBGs that LSBG\_exp+ser missed.}}

% the difference between LSBG_1comp and LSBG_2comps
\subsubsection{Comparison of LSBGs selected by one-component and two-component fitting}
%\textcolor{blue}{Because most of the galaxies in LSBG\_s\'{e}rsic are in the range of n < 1.5 (see Figure \ref{hist_n}), and the galaxies that perform two-component fitting are in the range of n > 1.5 in the single s\'{e}rsic fitting, there are very few duplicates between LSBG\_exp+deV, as well as LSBG\_exp+ser, and LSBG\_s\'{e}rsic.} 

For convenience, we will call the galaxies in LSBG\_exp and LSBG\_s\'{e}rsic ``LSBG\_1comp" (in total 1387 galaxies), and the galaxies in LSBG\_exp+deV and LSBG\_exp+ser ``LSBG\_2comps" (in total 247 galaxies). There are in total 1546 LSBGs with 88 duplicating in LSBG\_1comp and LSBG\_2comps, suggesting 15.98\% (247/1546) of our LSBGs hold bulge. It should be noted that this fraction might be smaller than it actually is, since we perform two-component fitting only for galaxies with $n > 1.5$ in the single s\'{e}rsic fitting, as a result, some galaxies with small bulges will be missed. Among the 88 overlapped galaxies of LSBG\_1comp and LSBG\_2comps, 96.59\% (85/88) of them have bulge-to-total ratio $\rm B/T < 0.25$, as blue histogram in Figure \ref{compare_1comp_2comps}(c) shown, suggesting that the 88 overlapped LSBGs have small bulges, they can be fitted well by a single component model. The remaining 1299 galaxies in LSBG\_1comp can be considered as a single disk without bulges. Even if they do contain a bulge, the B/T would be smaller than that of the overlaps.

\begin{figure}[h]
\centering
\includegraphics[width=8cm]{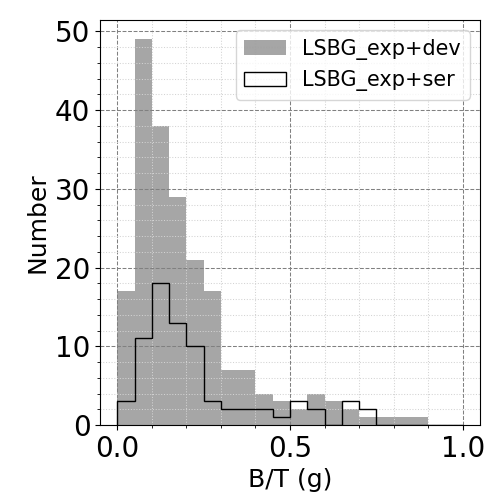}
\caption{The distribution of g band bulge-to-total ratio (B/T) of galaxies in LSBG\_exp+deV and LSBG\_exp+ser. Around 82.61\% of the galaxies in LSBG\_exp+deV and 77.33\% in LSBG\_exp+s\'{e}rsic have $\rm B/T < 0.3$.}
\label{hist_bt}
\end{figure}

The median values of $\rm R_{e,disk}$, g band stellar mass, H\Rmnum{1} mass, and W50 of LSBG\_2comps are 3.8 $\sim$ 4.1 kpc, 0.63 $\sim$ 0.76 dex, 0.44 $\sim$ 0.52 dex, and 30 $\sim$ 50 km/s larger than those of LSBG\_1comp, the median value of g band absolute magnitude of LSBG\_2comps is about 1.5 $\sim$ 1.7 mag brighter than that of LSBG\_1comp. Compared to LSBG\_1comp, the distributions of LSBG\_exp+deV and LSBG\_exp+ser have another peak at the red and massive (or bright) end, we point out the peaks in panels (c), (d), and (e) in Figure \ref{LSBG_all} by red arrows, the giant low surface brightness galaxies (gLSBGs) contribute a lot to this peak. However, the median value of g-r color distribution of LSBG\_2comps is only 0.03 $\sim$ 0.05 mag redder than that of LSBG\_1comp, and this is because the bulges are not prominent in the majority of our LSBG\_2comps. The distribution of B/T shown in Figure \ref{hist_bt} shows 82.61\% (171/207) of the galaxies in LSBG\_exp+deV and 77.33\% (58/75) the galaxies in LSBG\_exp+ser have $\rm B/T < 0.3$. 

\begin{figure}[h]
    \centering
    \includegraphics[width=8cm]{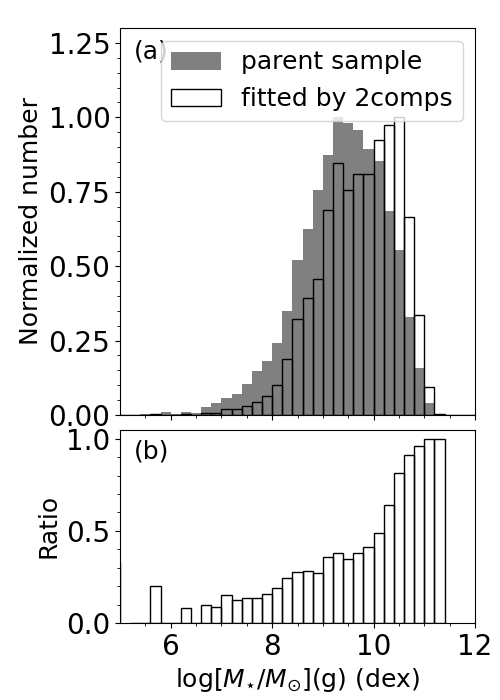}
    \caption{(a): The histograms of g band stellar mass of our parent sample and galaxies fitted by two-component models. The histogram is normalized by its peak value. (b): The label of y-axis is the ratio of galaxy number that fitted by two-component models and galaxy number of our parent sample.}
    \label{hist_stellar_mass}
\end{figure}

Figure \ref{hist_stellar_mass} shows the stellar mass distribution of our parent sample and galaxies that fitted by two-component models. The histogram is normalized by its peak value. We can see that the galaxies selected for two-component fitting are indeed the more massive galaxies in the parent sample, and the fraction of selected galaxies in each mass bin increases with the increase in stellar mass. This suggests that the result we obtained that the g-band stellar mass of the LSBG\_2comps is 0.63 $\sim$ 0.76 dex larger than that of the LSBG\_1comp is partly due to the selection effect. However, we think part of the result reflects the true differences of LSBG\_2comps and LSBG\_1comp: the disk component in LSBG\_2comps is larger than that in LSBG\_1comp (Figure \ref{LSBG_all}(b)), larger size often corresponds to larger mass. Moreover, the presence of bulge also contributes to the stellar mass observed in the LSBG\_2comps. 
%\textcolor{red}{\sout{This also suggests that LSBG\_2comps are not simply add a bulge to a low surface brightness disk, but a different kind of galaxy from LSBG\_1comp. The LSBG\_2comps is supplement to the LSBG\_1comp.}}

\subsubsection{Loss fraction of LSBGs with bulges}
\begin{figure}[h]
    \centering
    \includegraphics[width=\textwidth]{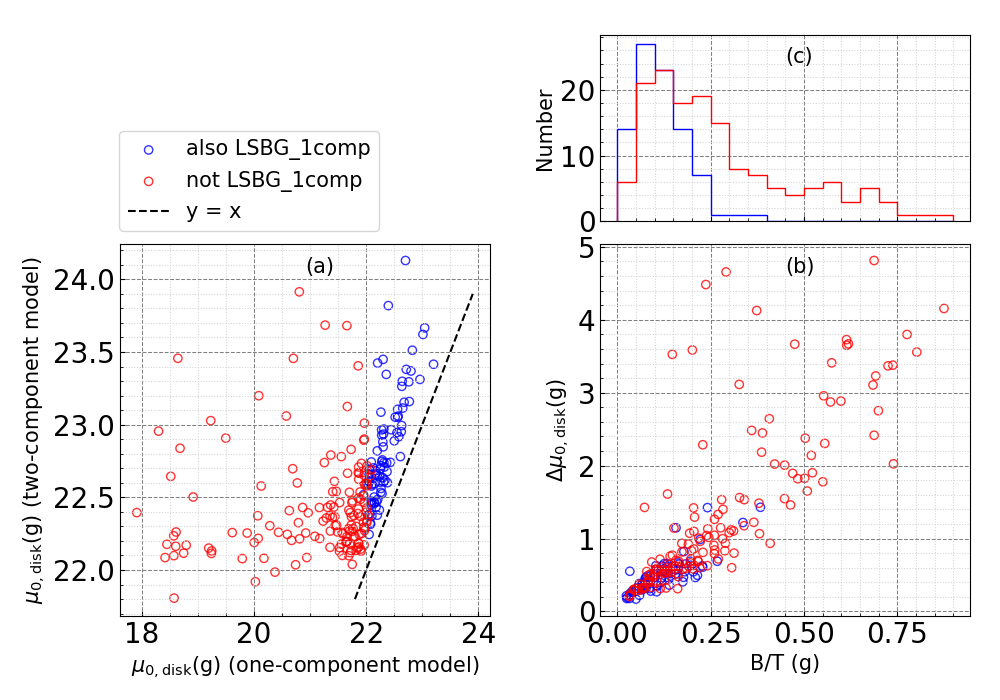}
    \caption{The difference of disk central surface brightness ($\rm \mu_{0,disk}$) derived from one-component model fitting and two-component model fitting. Among the 247 galaxies in LSBG\_2comps, 88 of them are identified as LSBGs by the single-component fitting (blue, denoted as ``also LSBG\_1comp"), and 151 galaxies are not (red, denoted as ``not LSBG\_1comp"). The black dashed line is the one-to-one line. All the $\mu_{0}$ values from two-component model fitting are dimmer than those from single-component model fitting, and the $\mu_{0}$ difference is related to the bulge-to-total ratio (B/T) of the galaxy. More than 60\% of our LSBG\_2comps will not be selected by a single-component fitting, especially for the galaxies with larger B/T.}
    \label{compare_1comp_2comps}
\end{figure}

Among our LSBG\_2comps, about 35.63\% (88/247) of them were also identified as LSBGs by single-component fitting (blue circles in Figure \ref{compare_1comp_2comps}, denoted as ``also LSBG\_1comp"), 61.13\% (151/247) were not identified as LSBGs by single-component fitting (red circles in Figure \ref{compare_1comp_2comps}, denoted as ``not LSBG\_1comp"), and 3.24\% (8/247) do not have good single-component fitting results because they were deleted in the second step of fitting results filtering (see Section \ref{Galfit fitting}). That is, more than 60\% of the LSBG\_2comps will not be selected if we adopt a single disk model only. The $\rm \mu_{0,disk}$ of two-component fitting is dimmer than that of single-component fitting, and the $\mu_{0}$ difference is related to the B/T of the galaxy. 72.47\% (179/247) of our LSBG\_2comps have $\rm B/T < 0.2$, and in this range the single-component fitting will result in an overestimation of the $\mu_{0}$ by approximately 0.8 $\rm mag\ arcsec^{-2}$.
%their b/a and PA are not consistent between g and r bands

\section{Giant low surface brightness galaxies} \label{gLSBGs}
Giant low surface brightness galaxies (gLSBGs) are a subset of LSBGs, they could have comparable or larger mass comparing with the Milky Way, but several times the size of the Milky Way. One of the most well-known gLSBG, Malin 1, was discovered by \cite{1987AJ.....94...23B}. Subsequent studies show Malin 1 has H\Rmnum{1} mass $M_{\rm H\Rmnum{1}} \sim 10^{11}M_{\odot}$ \citep{1987AJ.....94...23B}, disk central surface brightness $\rm \mu_{0,disk}(R)\ =\ 24.7\ mag\ arcsec^{-2}$ (for $h$ = 0.7; \citealt{2006PASA...23..165M}), and its optical diameter can reach 160 kpc \citep{2015ApJ...815L..29G}. UGC 1382 is also a very large gLSBG whose effective radius is 38 kpc \citep{2016ApJ...826..210H}. But gLSBGs whose size comparable with Malin 1 and UGC 1382 are rare, most of the gLSBGs have scale length from a few kpc to dozens of kpc (\citealt{1994AJ....107..530M, 1995AJ....109..558S}; \citetalias{2018MNRAS.476.4488H}; \citetalias{2023MNRAS.520L..85S}).

In Table \ref{gLSBG_standard}, we list different criteria used to select gLSBGs in previous works. In general, there are roughly three kinds of selection criteria. The first is the ``diffuseness index" criterion proposed by \cite{1995AJ....109..558S}: $\rm \mu_{0,B} + 5\times log10(R_{s, B}) > 27.0$ where the scale length is in unit of $h^{-1}$ kpc. The second is the mass criterion: \cite{2018MNRAS.476.4488H} (hereafter \citetalias{2018MNRAS.476.4488H}) selected 41 gLSBGs from their spiral LSBGs with the criteria $M_{\rm H\Rmnum{1}} > 10^{9}M_{\odot}$ and $M_{\star} > 10^{10}M_{\odot}$; \cite{2023AJ....165..263O} adopted $M_{\rm H\Rmnum{1}} \geqslant 10^{10}M_{\odot}$ to identify massive LSBGs. The third is the size criterion: \cite{1997ARAA...35..267I} claimed that gLSBGs have scale lengths in excess of 10 kpc; \cite{2023MNRAS.520L..85S} (hereafter \citetalias{2023MNRAS.520L..85S}) detected 42 gLSBGs with $\rm R_{iso, 28B} > 50\ kpc$ or four times scale-length $\rm 4R_{s} > 50\ kpc$. 

\begin{table}[h]
\centering
\caption{Criteria for gLSBG selection.} \label{gLSBG_standard}
\begin{tabular}{cc}
\hline
Criterion & Reference \\
\hline
$\rm \mu_{0,B} + 5\times log10(R_{s, B}) > 27.0$                          & \cite{1995AJ....109..558S}       \\
$M_{\rm H\Rmnum{1}} > 10^{9}M_{\odot}$ and $M_{\star} > 10^{10}M_{\odot}$ & \cite{2018MNRAS.476.4488H} \\
$M_{\rm H\Rmnum{1}} \geq 10^{10}M_{\odot}$                                & \cite{2023AJ....165..263O}       \\
$\rm R_{s} > 10\ kpc$                                                     & \cite{1997ARAA...35..267I}       \\ 
$\rm R_{iso, 28B} > 50\ kpc$ or $\rm 4R_{s} > 50\ kpc$                          & \cite{2023MNRAS.520L..85S} \\
\hline
\end{tabular}
\end{table}

\begin{figure}[h]
    \centering
    \includegraphics[width=8cm]{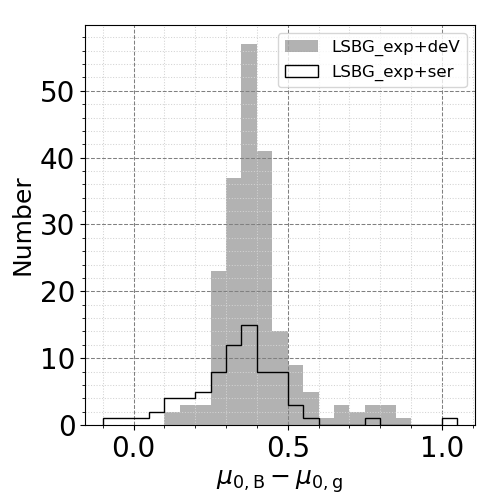}
    \caption{The histogram of the $\mu_{0}$ difference between B and g band for LSBG\_exp+deV and LSBG\_exp+ser samples.}
    \label{mu0_B_g}
\end{figure}

\begin{figure}[h]
    \centering
    \includegraphics[width=\textwidth]{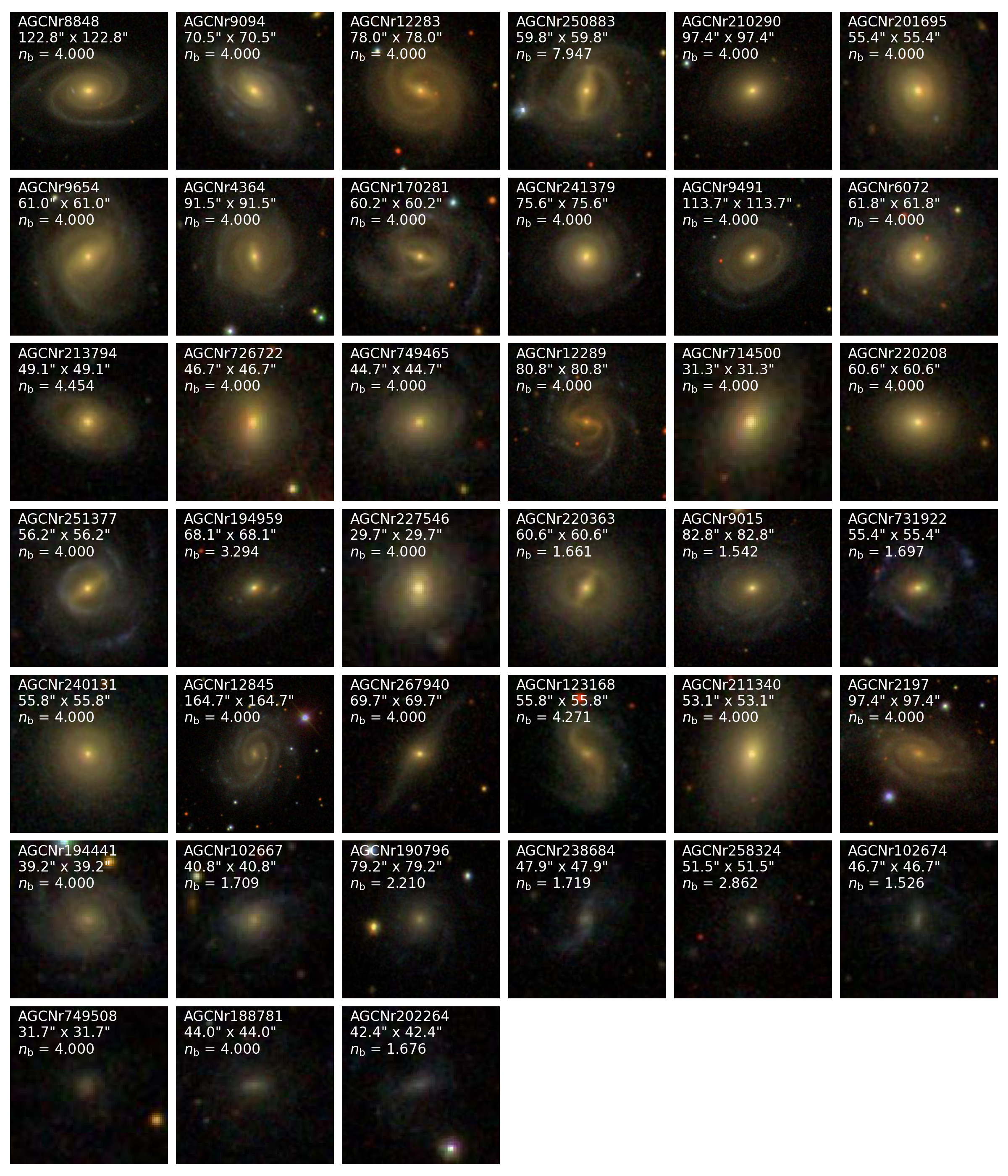}
    \caption{The colored images from SDSS of the gLSBGs. The first 31 galaxies are the gLSBGs whose stellar mass is larger than $10^{10} M_{\odot}$, and the last 8 galaxies are the gLSBGs whose stellar mass is smaller than $10^{10} M_{\odot}$. The last 8 galaxies seem different from the gLSBGs that most people know, more like normal LSBGs with bulge. We will eliminate these 8 galaxies from our gLSBGs sample.}
    \label{images_gLSBG}
\end{figure}

This work will use these three kinds of criteria to select gLSBGs from our LSBG\_2comps. Figure \ref{mu0_B_g} shows that the $\mu_{0}$ difference between B and g bands in our sample is $\rm \sim 0.35\ mag\ arcsec^{-2}$, so we adopt ``diffuseness index" $\rm \mu_{0,g} + 5\times log10(R_{s, g}) > 26.65$ and select a sample named ``gLSBG\_diffuse" which contains 7 gLSBGs from LSBG\_exp+deV and 8 gLSBGs from LSBG\_exp+ser with 2 galaxies replicated. Following the mass criterion in \citetalias{2018MNRAS.476.4488H}, we select a sample named ``gLSBG\_mass" which contains 27 gLSBGs from LSBG\_exp+deV and 7 gLSBGs from LSBG\_exp+ser with 3 galaxies replicated. We also select a sample named ``gLSBG\_size" under the criterion $\rm R_{s, g} \geq 10\ kpc$ which contains 3 galaxies from LSBG\_exp+deV and 2 gLSBGs from LSBG\_exp+ser with no galaxy replicated. There are totally 39 non-repetitive gLSBGs, the detailed information is listed in Table \ref{table_gLSBG}, and the SDSS color images of them are shown in Figure \ref{images_gLSBG}.

\begin{table}[h]
\centering
\caption{Parameters of gLSBGs.} \label{table_gLSBG}
\resizebox{\textwidth}{!}{
\begin{threeparttable}
\begin{tabular}{ccccccccccc}
\hline
AGCNr & RA & DEC & g-r & $M_{\rm g}$ & log(HI) & log($M_{\star}$) & $\rm R_{s, g}$ & $\mu_{0, g}$ & $n_{\rm b, g}$ & criterion \\
 & (J2000) & (J2000) & (mag) & (mag) & ($M_{\odot}$) & ($M_{\odot}$) & (kpc) & ($\rm mag\ arcsec^{-2}$) &  &  \\
\hline
  8848 & 209.014 & 13.506 & 0.743 & -21.421 & 10.130 & 10.912 & 10.088 & 22.235 & 4.000 & mass, size; exp+deV \\ \hline
  9094 & 213.121 & 24.636 & 0.650 & -21.628 & 10.640 & 10.851 &  9.486 & 22.114 & 4.000 & mass; exp+deV \\ \hline
 12283 & 344.832 & 24.107 & 0.700 & -21.426 &  9.810 & 10.847 &  8.643 & 22.086 & 4.000 & mass; exp+deV \\ \hline
250883 & 232.229 &  7.745 & 0.698 & -21.397 &  9.950 & 10.833 &  9.352 & 22.189 & 7.947 & mass; exp+ser \\ \hline
210290 & 170.720 & 27.584 & 0.761 & -21.085 & 10.000 & 10.807 & 11.609 & 23.457 & 4.000 & diffuse, mass, size; exp+deV \\ \hline
201695 & 164.286 & 14.925 & 0.781 & -20.837 & 10.150 & 10.738 &  6.611 & 22.131 & 4.000 &    mass; exp+deV \\ \hline
  9654 & 225.236 & 11.517 & 0.720 & -21.032 &  9.980 & 10.721 &  6.841 & 21.919 & 4.000 &    mass; exp+deV \\ \hline
  4364 & 125.603 & 25.509 & 0.656 & -21.252 &  9.990 & 10.709 &  9.041 & 22.078 & 4.000 &    mass; exp+deV \\ \hline
170281 & 119.332 & 11.206 & 0.627 & -21.261 & 10.090 & 10.669 &  9.840 & 22.087 & 4.000 &    mass; exp+deV \\ \hline
241379 & 211.430 & 25.231 & 0.727 & -20.866 &  9.940 & 10.665 &  4.777 & 22.171 & 4.000 &    mass; exp+deV \\ \hline
  9491 & 221.061 &  4.219 & 0.656 & -21.053 & 10.030 & 10.630 & 10.374 & 22.395 & 4.000 & diffuse, mass, size; exp+deV \\ \hline
  6072 & 164.935 & 10.071 & 0.610 & -21.200 & 10.420 & 10.618 &  7.825 & 22.373 & 4.000 &    mass; exp+deV \\ \hline
213794 & 168.103 &  7.495 & 0.744 & -20.579 & 10.280 & 10.578 &  8.209 & 22.151 & 4.454 &    mass; exp+ser \\ \hline
726722 & 218.947 & 24.753 & 0.780 & -20.435 & 10.350 & 10.577 &  5.667 & 23.028 & 4.000 &    mass; exp+deV \\ \hline
749465 & 207.177 & 25.757 & 0.618 & -20.958 & 10.440 & 10.533 &  8.160 & 22.305 & 4.000 &    mass; exp+deV \\ \hline
 12289 & 344.923 & 24.075 & 0.572 & -21.127 & 10.300 & 10.528 &  9.133 & 22.709 & 4.000 & diffuse, mass; exp+deV \\ \hline
714500 & 225.195 &  8.144 & 0.739 & -20.420 & 10.070 & 10.507 &  5.231 & 22.177 & 4.000 &    mass; exp+deV \\ \hline
220208 & 183.279 & 16.086 & 0.767 & -20.267 &  9.640 & 10.488 &  3.396 & 22.163 & 4.000 &    mass; exp+deV \\ \hline
251377 & 239.508 & 14.964 & 0.559 & -20.926 & 10.400 & 10.428 &  7.391 & 22.036 & 4.000 &    mass; exp+deV \\ \hline
194959 & 141.762 & 27.845 & 0.652 & -20.340 & 10.190 & 10.339 & 10.657 & 22.956 & 3.294 & diffuse, mass, size; exp+ser \\ \hline
227546 & 192.794 & 26.795 & 0.668 & -20.260 & 10.310 & 10.332 &  4.788 & 22.261 & 4.000 &    mass; exp+deV \\ \hline
220363 & 184.851 & 12.301 & 0.642 & -20.337 & 9.680 & 10.321  &  6.062 & 22.145 & 4.000 &    mass; exp+deV \\
---    & ---     & ---    & ---   & ---     & ---   & ---     &  5.762 & 21.986 & 1.661 &    mass; exp+ser \\ \hline
  9015 & 211.485 &  9.026 & 0.582 & -20.421 & 10.020 & 10.261 &  7.884 & 22.442 & 4.000 &    mass; exp+deV \\
---    & ---     & ---    & ---   & ---     & ---   & ---     &  6.913 & 22.099 & 1.542 &    mass; exp+ser \\ \hline
731922 & 181.449 & 24.116 & 0.496 & -20.674 & 10.370 & 10.230 & 10.026 & 22.502 & 1.697 & diffuse, mass, size; exp+ser \\ \hline
240131 & 212.341 &  8.907 & 0.631 & -20.099 &  9.760 & 10.210 &  4.699 & 22.260 & 4.000 &    mass; exp+deV \\ \hline
 12845 & 358.924 & 31.900 & 0.562 & -20.355 & 10.180 & 10.205 &  9.202 & 22.433 & 4.000 &    mass; exp+deV \\ \hline
267940 & 240.576 & 12.274 & 0.710 & -19.749 & 10.180 & 10.193 &  8.557 & 22.644 & 4.000 &    mass; exp+deV \\ \hline
123168 &  40.333 & 29.147 & 0.524 & -20.368 &  9.920 & 10.151 &  6.708 & 22.093 & 4.000 &    mass; exp+deV \\ 
---    & ---     & ---    & ---   & ---     & ---    & ---    &  6.720 & 22.098 & 4.271 &    mass; exp+ser \\ \hline
211340 & 177.467 &  6.668 & 0.703 & -19.664 &  9.350 & 10.147 &  3.238 & 22.838 & 4.000 &    mass; exp+deV \\ \hline
  2197 &  40.858 & 31.471 & 0.525 & -20.342 &  9.920 & 10.141 &  5.520 & 22.039 & 4.000 &    mass; exp+deV \\ \hline
194441 & 137.940 & 27.659 & 0.487 & -20.249 &  9.870 & 10.044 &  6.336 & 22.134 & 4.000 &    mass; exp+deV \\ \hline
102667 &  11.070 & 25.766 & 0.403 & -19.598 &  9.970 &  9.655 &  6.255 & 23.685 & 1.709 & diffuse; exp+ser \\ \hline
190796 & 137.776 & 13.122 & 0.340 & -19.836 & 10.140 &  9.651 &  8.407 & 23.199 & 2.210 & diffuse; exp+ser \\ \hline
238684 & 195.473 &  5.387 & 0.304 & -19.316 &  9.990 &  9.387 &  8.255 & 23.086 & 1.719 & diffuse; exp+ser \\ \hline
258324 & 234.036 &  5.793 & 0.343 & -18.765 &  9.720 &  9.227 &  7.103 & 23.718 & 4.000 & diffuse; exp+deV \\ 
---    & ---     & ---    & ---   & ---     & ---    & ---    &  7.166 & 23.381 & 2.862 & diffuse; exp+ser \\ \hline
102674 &  12.310 & 25.294 & 0.219 & -19.168 & 10.020 &  9.195 &  7.633 & 23.566 & 4.000 & diffuse; exp+deV \\ 
---    & ---     & ---    & ---   & ---     & ---    &  ---   &  7.316 & 23.264 & 1.526 & diffuse; exp+ser \\ \hline
749508 & 244.915 & 25.563 & 0.346 & -18.541 &  9.820 &  9.143 &  8.117 & 24.429 & 4.000 & diffuse; exp+deV \\ \hline
188781 & 132.116 & 16.192 & 0.397 & -17.625 &  9.460 &  8.855 &  4.724 & 24.130 & 4.000 & diffuse; exp+deV \\ \hline
202264 & 152.444 & 12.082 & 0.268 & -17.802 &  9.650 &  8.726 &  6.468 & 23.819 & 1.676 & diffuse; exp+ser \\ \hline
\end{tabular}
\begin{tablenotes}
\footnotesize
\item \normalsize The table is sorted by the g band stellar mass (log($M_{\star}$)). The last 8 galaxies have stellar mass smaller than $10^{10}M_{\odot}$, and all of them are selected by the ``diffuseness index" criterion. They seem different from the gLSBGs that most people know, more like normal LSBGs with bulge. We will eliminate these 8 galaxies from our gLSBGs sample.
\end{tablenotes}
\end{threeparttable}
}
\end{table}

\begin{figure}[h]
    \centering
    \includegraphics[width=\textwidth]{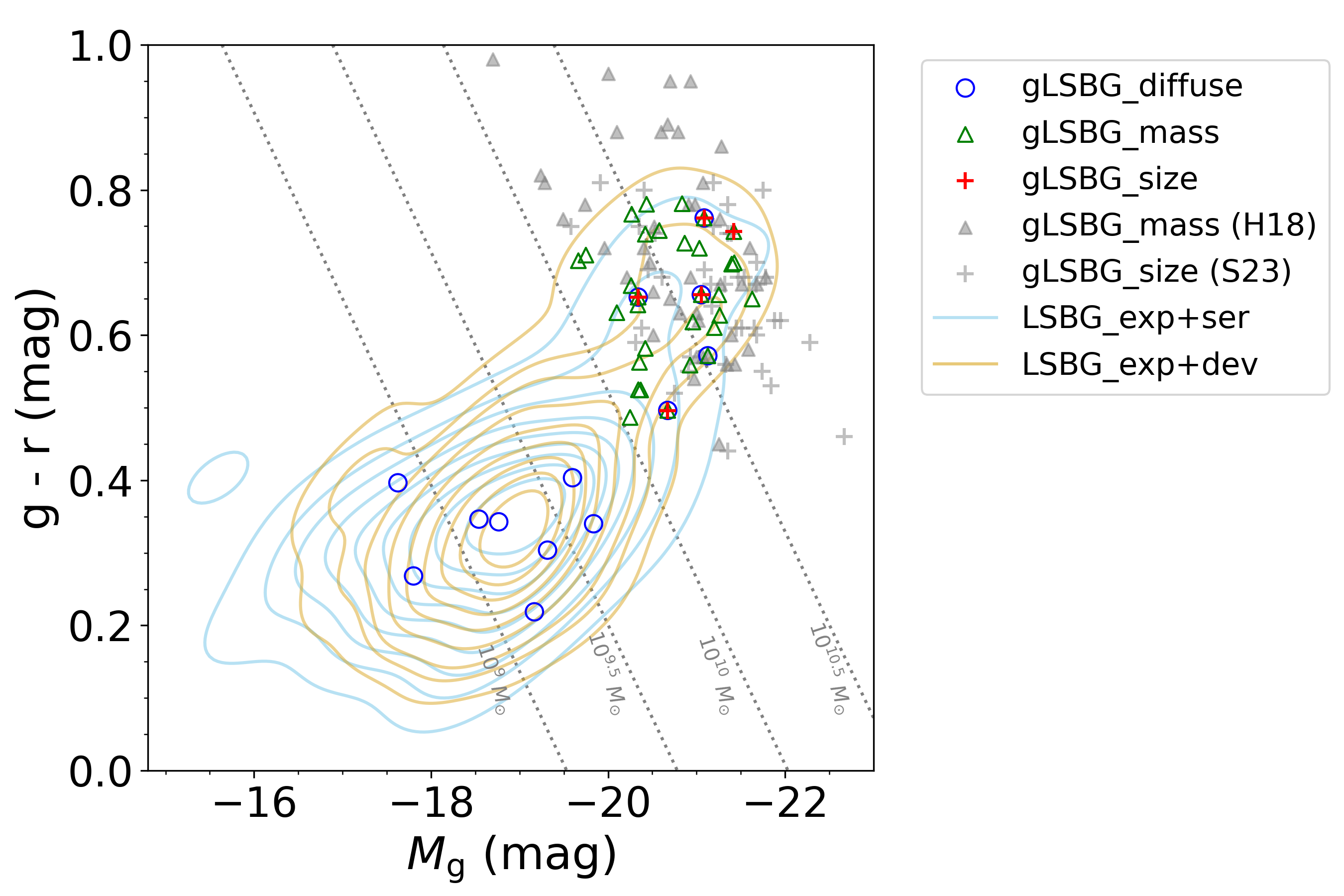}
    \caption{The color-magnitude diagram of gLSBGs superimposed on kernel density estimation (KDE) of LSBG\_exp+deV (light yellow) and LSBG\_exp+ser (light blue). Different shapes of marker represent the gLSBGs selected from different criteria. The gray triangles and gray crosses represent gLSBGs from \citetalias{2018MNRAS.476.4488H} and \citetalias{2023MNRAS.520L..85S}, respectively. The gray dotted lines represent specific stellar mass values.}
    \label{CMR_gLSBG}
\end{figure}

Figure \ref{CMR_gLSBG} shows the color-magnitude diagram of gLSBGs superimposed on kernel density estimation (KDE) of LSBG\_exp+deV (light yellow) and LSBG\_exp+ser (light blue). The grey triangles and gray crosses represent gLSBGs from \citetalias{2018MNRAS.476.4488H} and \citetalias{2023MNRAS.520L..85S}, respectively. The gray dotted lines represent stellar mass of $10^{9}M_{\odot}$, $10^{9.5}M_{\odot}$, $10^{10}M_{\odot}$, and $10^{10.5}M_{\odot}$. Different shapes of marker represent the gLSBGs selected from different criteria. Among our 39 non-repetitive gLSBGs, 31 of them lie on the right side of the $10^{10}M_{\odot}$ stellar mass line. 90.32\% (28/31) out of the 31 gLSBGs have the s\'{e}rsic $n$ index of the bulge component $n_{\rm b} > 2$, indicating most of our gLSBGs hold a classical bulge. These 31 gLSBGs locate at the same region as gLSBGs in \citetalias{2018MNRAS.476.4488H} and \citetalias{2023MNRAS.520L..85S}, they also seem similar to the gLSBGs that most people recognized which are red in color and giant in luminosity, mass, and size. There are 8 galaxies on the left side of the $10^{10}M_{\odot}$ stellar mass line, and all these gLSBGs are selected by the ``diffuseness index" criterion. They seem different from the gLSBGs that most people know, more like normal LSBGs with bulge. This is because the dim $\rm \mu_{0,disk}$ contributes a lot to the ``diffuseness index". The median value of $\rm \mu_{0,disk}$ of these 8 gLSBGs is $\rm \sim 1.356\ mag\ arcsec^{-2}$ dimmer than that of the 31 gLSBGs whose stellar mass is larger than $10^{10}M_{\odot}$. This is the drawback of the ``diffuseness index" criterion. We will eliminate these 8 galaxies from our gLSBGs sample. Our result suggests that for gas-rich LSBGs, $M_{\star} > 10^{10}M_{\odot}$ is a good criterion to distinguish between gLSBGs and normal LSBGs with bulge.

\section{Summary} \label{summary}
In this work, we carried out fitting of four models, exponential, s\'{e}rsic, exp+deV, and exp+ser, to check the systematic effect in selecting LSBGs. Our parent sample is all the galaxies in $\upalpha$.40 SDSS DR7 sample, and the images we used are the SDSS images after background reestimation (\citetalias{2015AJ....149..199D}). We took photometry and two-dimensional fitting for these images.

According to the criteria that the B band central surface brightness of the disk component $\rm \mu_{0,disk}(B) \geqslant 22.5\ mag\ arcsec^{-2}$ and the axis-ratio of the disk component $\rm b/a > 0.3$, we selected 1105, 1038, 207, and 75 none-edge-on LSBGs from each of the models. Because our parent sample is rich in H\Rmnum{1} gas, most of our LSBGs are blue. The LSBGs selected by s\'{e}rsic fitting are in principle the same type of galaxies as exponential fitting. LSBG\_exp and LSBG\_s\'{e}rsic have 756 galaxies (68.42\% of LSBG\_exp and 72.83\% of LSBG\_s\'{e}rsic) in common, the rest of the discrepancy is due to the difference between the exponential and s\'{e}rsic models in obtaining $\mu_{0}$. When $0.5 \leqslant n \leqslant 1.5$, the relation of $\mu_{0}$ obtained from s\'{e}rsic and exponential models can be written as ${\rm \mu_{0,s\acute{e}rsic} - \mu_{0,exp}} = -1.34(n-1)$. 
There are 1546 non-repetitive LSBGs, at least 15.89\% of them hold a bulge. The galaxies in LSBG\_2comps have larger disk, luminosity, and mass than galaxies in LSBG\_1comp. But the bulges is not prominent in the majority of our LSBGs. More than 60\% of the galaxies in LSBG\_2comps will not be selected if we adopt a single disk model only. We also identified 31 non-repetitive gLSBGs, with 90.32\% of them hold a classical bulge. They are located at the same region in the color-magnitude diagram as normal gLSBGs in literature. Based on our gas-rich LSBGs, we find $M_{\star} > 10^{10}M_{\odot}$ is a good criterion to distinguish between gLSBGs and normal LSBGs with bulge after we compared three kinds of criteria.

\normalem
\begin{acknowledgements}
This work is supported by the National Key R\&D Program of China (grant No.2022YFA1602901). We acknowledge the support of the National Natural Science Foundation of China (NSFC) grant Nos.12090040, 12090041, and 12003043. D.W. is supported by the Youth Innovation Promotion Association, CAS (No.2020057) and the science research grants of CSST from the China Manned Space Project. We acknowledge the support of the National Natural Science Foundation of China (NSFC) grant Nos.11733006 and U1931109. This work is supported by the Strategic Priority Research Program of the Chinese Academy of Sciences, Grant No. XDB0550100. We acknowledge useful discussions with Mr. Zehao Zhong and Mr. Yu Zhang. This work was partially supported by the Open Project Program of the Key Laboratory of Optical Astronomy, National Astronomical Observatories, Chinese Academy of Sciences.

\end{acknowledgements}
  
\bibliographystyle{raa}
\bibliography{lsbg}

\begin{thebibliography}{60}
\providecommand\natexlab[1]{#1}
\providecommand\JournalTitle[1]{#1}

\bibitem[{Abazajian} {et~al.}(2009)]{2009ApJS..182..543A}
{Abazajian}, K.~N., {Adelman-McCarthy}, J.~K., {Ag{\"u}eros}, M.~A., {et~al.}
  2009, \apjs, 182, 543

\bibitem[{Adelman-McCarthy} {et~al.}(2006)]{2006ApJS..162...38A}
{Adelman-McCarthy}, J.~K., {Ag{\"u}eros}, M.~A., {Allam}, S.~S., {et~al.} 2006,
  \apjs, 162, 38

\bibitem[{Adelman-McCarthy} {et~al.}(2008)]{2008ApJS..175..297A}
{Adelman-McCarthy}, J.~K., {Ag{\"u}eros}, M.~A., {Allam}, S.~S., {et~al.} 2008,
  \apjs, 175, 297

\bibitem[{Bertin} \& {Arnouts}(1996)]{1996A&AS..117..393B}
{Bertin}, E., \& {Arnouts}, S. 1996, \aaps, 117, 393

\bibitem[{Bothun} {et~al.}(1987)]{1987AJ.....94...23B}
{Bothun}, G.~D., {Impey}, C.~D., {Malin}, D.~F., \& {Mould}, J.~R. 1987, \aj,
  94, 23

\bibitem[{Brown} {et~al.}(2001)]{2001AJ....122..714B}
{Brown}, W.~R., {Geller}, M.~J., {Fabricant}, D.~G., \& {Kurtz}, M.~J. 2001,
  \aj, 122, 714

\bibitem[{Burkholder} {et~al.}(2001)]{2001AJ....122.2318B}
{Burkholder}, V., {Impey}, C., \& {Sprayberry}, D. 2001, \aj, 122, 2318

\bibitem[{Cao} {et~al.}(2017)]{2017AJ....154..116C}
{Cao}, T.-W., {Wu}, H., {Du}, W., {et~al.} 2017, \aj, 154, 116

\bibitem[{Capaccioli}(1989)]{1989woga.conf..208C}
{Capaccioli}, M. 1989, in World of Galaxies (Le Monde des Galaxies), ed.
  J.~{Corwin}, Harold~G. \& L.~{Bottinelli} (Springer New York, NY), 208

\bibitem[{Courteau}(1996)]{1996ApJS..103..363C}
{Courteau}, S. 1996, \apjs, 103, 363

\bibitem[{de Blok} {et~al.}(1996)]{1996MNRAS.283...18D}
{de Blok}, W.~J.~G., {McGaugh}, S.~S., \& {van der Hulst}, J.~M. 1996, \mnras,
  283, 18

\bibitem[{de Blok} {et~al.}(1995)]{1995MNRAS.274..235D}
{de Blok}, W.~J.~G., {van der Hulst}, J.~M., \& {Bothun}, G.~D. 1995, \mnras,
  274, 235

\bibitem[{Dey} {et~al.}(2019)]{2019AJ....157..168D}
{Dey}, A., {Schlegel}, D.~J., {Lang}, D., {et~al.} 2019, \aj, 157, 168

\bibitem[{Du} {et~al.}(2020)]{2020AJ....159..138D}
{Du}, W., {Cheng}, C., {Zheng}, Z., \& {Wu}, H. 2020, \aj, 159, 138

\bibitem[{Du} {et~al.}(2015)]{2015AJ....149..199D}
{Du}, W., {Wu}, H., {Lam}, M.~I., {et~al.} 2015, \aj, 149, 199

\bibitem[{Fisher} \& {Drory}(2008)]{2008AJ....136..773F}
{Fisher}, D.~B., \& {Drory}, N. 2008, \aj, 136, 773

\bibitem[{Galaz} {et~al.}(2011)]{2011ApJ...728...74G}
{Galaz}, G., {Herrera-Camus}, R., {Garcia-Lambas}, D., \& {Padilla}, N. 2011,
  \apj, 728, 74

\bibitem[{Galaz} {et~al.}(2015)]{2015ApJ...815L..29G}
{Galaz}, G., {Milovic}, C., {Suc}, V., {et~al.} 2015, \apjl, 815, L29

\bibitem[{Giovanelli} {et~al.}(2005)]{2005AJ....130.2598G}
{Giovanelli}, R., {Haynes}, M.~P., {Kent}, B.~R., {et~al.} 2005, \aj, 130, 2598

\bibitem[{Greco} {et~al.}(2018)]{2018ApJ...857..104G}
{Greco}, J.~P., {Greene}, J.~E., {Strauss}, M.~A., {et~al.} 2018, \apj, 857,
  104

\bibitem[{Hagen} {et~al.}(2016)]{2016ApJ...826..210H}
{Hagen}, L. M.~Z., {Seibert}, M., {Hagen}, A., {et~al.} 2016, \apj, 826, 210

\bibitem[{Haynes} {et~al.}(2011)]{2011AJ....142..170H}
{Haynes}, M.~P., {Giovanelli}, R., {Martin}, A.~M., {et~al.} 2011, \aj, 142,
  170

\bibitem[{He} {et~al.}(2020)]{2020ApJS..248...33H}
{He}, M., {Wu}, H., {Du}, W., {et~al.} 2020, \apjs, 248, 33

\bibitem[{He} {et~al.}(2013)]{2013ApJ...773...37H}
{He}, Y.~Q., {Xia}, X.~Y., {Hao}, C.~N., {et~al.} 2013, \apj, 773, 37

\bibitem[{Honey} {et~al.}(2018)]{2018MNRAS.476.4488H}
{Honey}, M., {van Driel}, W., {Das}, M., \& {Martin}, J.~M. 2018, \mnras, 476,
  4488

\bibitem[{Huang} {et~al.}(2014)]{2014ApJ...793...40H}
{Huang}, S., {Haynes}, M.~P., {Giovanelli}, R., {et~al.} 2014, \apj, 793, 40

\bibitem[{Hyde} \& {Bernardi}(2009)]{2009MNRAS.394.1978H}
{Hyde}, J.~B., \& {Bernardi}, M. 2009, \mnras, 394, 1978

\bibitem[{Impey} \& {Bothun}(1997)]{1997ARAA...35..267I}
{Impey}, C., \& {Bothun}, G. 1997, \araa, 35, 267

\bibitem[{Kron}(1980)]{1980ApJS...43..305K}
{Kron}, R.~G. 1980, \apjs, 43, 305

\bibitem[{Lauer} {et~al.}(2007)]{2007ApJ...662..808L}
{Lauer}, T.~R., {Faber}, S.~M., {Richstone}, D., {et~al.} 2007, \apj, 662, 808

\bibitem[{Lei} {et~al.}(2019)]{2019ApJS..242...11L}
{Lei}, F.-J., {Wu}, H., {Zhu}, Y.-N., {et~al.} 2019, \apjs, 242, 11

\bibitem[{Lei} {et~al.}(2018)]{2018ApJS..235...18L}
{Lei}, F.-J., {Wu}, H., {Du}, W., {et~al.} 2018, \apjs, 235, 18

\bibitem[{Liang} {et~al.}(2010)]{2010MNRAS.409..213L}
{Liang}, Y.~C., {Zhong}, G.~H., {Hammer}, F., {et~al.} 2010, \mnras, 409, 213

\bibitem[{Liu} {et~al.}(2008)]{2008MNRAS.385...23L}
{Liu}, F.~S., {Xia}, X.~Y., {Mao}, S., {Wu}, H., \& {Deng}, Z.~G. 2008, \mnras,
  385, 23

\bibitem[{Martin} {et~al.}(2019)]{2019MNRAS.485..796M}
{Martin}, G., {Kaviraj}, S., {Laigle}, C., {et~al.} 2019, \mnras, 485, 796

\bibitem[{McGaugh}(1996)]{1996IAUS..171...97M}
{McGaugh}, S. 1996, in New Light on Galaxy Evolution, ed. R.~{Bender} \& R.~L.
  {Davies}, Vol. 171 (Springer Dordrecht), 97

\bibitem[{McGaugh} \& {Bothun}(1994)]{1994AJ....107..530M}
{McGaugh}, S.~S., \& {Bothun}, G.~D. 1994, \aj, 107, 530

\bibitem[{McGaugh} {et~al.}(1995{\natexlab{a}})]{1995AJ....110..573M}
{McGaugh}, S.~S., {Bothun}, G.~D., \& {Schombert}, J.~M. 1995{\natexlab{a}},
  \aj, 110, 573

\bibitem[{McGaugh} {et~al.}(1995{\natexlab{b}})]{1995AJ....109.2019M}
{McGaugh}, S.~S., {Schombert}, J.~M., \& {Bothun}, G.~D. 1995{\natexlab{b}},
  \aj, 109, 2019

\bibitem[{Moore} \& {Parker}(2006)]{2006PASA...23..165M}
{Moore}, L., \& {Parker}, Q.~A. 2006, \pasa, 23, 165

\bibitem[{O'Neil}(2004)]{2004AJ....128.2080O}
{O'Neil}, K. 2004, \aj, 128, 2080

\bibitem[{O'Neil} {et~al.}(1997)]{1997AJ....113.1212O}
{O'Neil}, K., {Bothun}, G.~D., \& {Cornell}, M.~E. 1997, \aj, 113, 1212

\bibitem[{O'Neil} {et~al.}(2003)]{2003ApJ...588..230O}
{O'Neil}, K., {Schinnerer}, E., \& {Hofner}, P. 2003, \apj, 588, 230

\bibitem[{O'Neil} {et~al.}(2023)]{2023AJ....165..263O}
{O'Neil}, K., {Schneider}, S.~E., {van Driel}, W., {et~al.} 2023, \aj, 165, 263

\bibitem[{Pahwa} \& {Saha}(2018)]{2018MNRAS.478.4657P}
{Pahwa}, I., \& {Saha}, K. 2018, \mnras, 478, 4657

\bibitem[{Peng} {et~al.}(2002)]{2002AJ....124..266P}
{Peng}, C.~Y., {Ho}, L.~C., {Impey}, C.~D., \& {Rix}, H.-W. 2002, \aj, 124, 266

\bibitem[{Peng} {et~al.}(2010)]{2010AJ....139.2097P}
{Peng}, C.~Y., {Ho}, L.~C., {Impey}, C.~D., \& {Rix}, H.-W. 2010, \aj, 139,
  2097

\bibitem[{Pickering} {et~al.}(1997)]{1997AJ....114.1858P}
{Pickering}, T.~E., {Impey}, C.~D., {van Gorkom}, J.~H., \& {Bothun}, G.~D.
  1997, \aj, 114, 1858

\bibitem[{Pizzella} {et~al.}(2008)]{2008MNRAS.387.1099P}
{Pizzella}, A., {Corsini}, E.~M., {Sarzi}, M., {et~al.} 2008, \mnras, 387, 1099

\bibitem[{Rosenbaum} {et~al.}(2009)]{2009A&A...504..807R}
{Rosenbaum}, S.~D., {Krusch}, E., {Bomans}, D.~J., \& {Dettmar}, R.~J. 2009,
  \aap, 504, 807

\bibitem[{Saburova} {et~al.}(2023)]{2023MNRAS.520L..85S}
{Saburova}, A.~S., {Chilingarian}, I.~V., {Kulier}, A., {et~al.} 2023, \mnras,
  520, L85

\bibitem[{Simard} {et~al.}(2011)]{2011ApJS..196...11S}
{Simard}, L., {Mendel}, J.~T., {Patton}, D.~R., {Ellison}, S.~L., \&
  {McConnachie}, A.~W. 2011, \apjs, 196, 11

\bibitem[{Smith} {et~al.}(2002)]{2002AJ....123.2121S}
{Smith}, J.~A., {Tucker}, D.~L., {Kent}, S., {et~al.} 2002, \aj, 123, 2121

\bibitem[{Sprayberry} {et~al.}(1995)]{1995AJ....109..558S}
{Sprayberry}, D., {Impey}, C.~D., {Bothun}, G.~D., \& {Irwin}, M.~J. 1995, \aj,
  109, 558

\bibitem[{Tanoglidis} {et~al.}(2021)]{2021ApJS..252...18T}
{Tanoglidis}, D., {Drlica-Wagner}, A., {Wei}, K., {et~al.} 2021, \apjs, 252, 18

\bibitem[{Trujillo} {et~al.}(2001)]{2001MNRAS.321..269T}
{Trujillo}, I., {Aguerri}, J.~A.~L., {Cepa}, J., \& {Guti{\'e}rrez}, C.~M.
  2001, \mnras, 321, 269

\bibitem[{van der Hulst} {et~al.}(1993)]{1993AJ....106..548V}
{van der Hulst}, J.~M., {Skillman}, E.~D., {Smith}, T.~R., {et~al.} 1993, \aj,
  106, 548

\bibitem[{Willmer}(2018)]{2018ApJS..236...47W}
{Willmer}, C. N.~A. 2018, \apjs, 236, 47

\bibitem[{York} {et~al.}(2000)]{2000AJ....120.1579Y}
{York}, D.~G., {Adelman}, J., {Anderson}, John~E., J., {et~al.} 2000, \aj, 120,
  1579

\bibitem[{Zhong} {et~al.}(2008)]{2008MNRAS.391..986Z}
{Zhong}, G.~H., {Liang}, Y.~C., {Liu}, F.~S., {et~al.} 2008, \mnras, 391, 986

\end{thebibliography}

\end{document}